\documentstyle[preprint,revtex]{aps}

\begin{document}
\draft

\begin{title}
Study of the magnetic susceptibility in the spin-Peierls system \\
CuGeO$_3$
\end{title}

\author{Jos\'e Riera and Ariel Dobry}

\begin{instit}
Instituto de F\'{\i}sica Rosario (C.O.N.I.C.E.T.) y
Departamento de F\'{\i}sica,    \\
Universidad Nacional de Rosario,
Av. Pellegrini 250,  2000-Rosario, Argentina.
\end{instit}

\receipt{}

\begin{abstract}
We study numerically, using a one-dimensional Heisenberg model,
the spin-Peierls transition in the linear Cu$^{2+}$ spin-1/2
chains in the inorganic compound CuGeO$_3$ which has been recently
observed experimentally. We suggest that the magnetic susceptibility,
the temperature dependence of the spin gap and the spin-Peierls
transition temperature of this material can be reasonably described
by including nearest and next nearest neighbor antiferromagnetic
interactions along the chain.  We estimate that the nearest neighbor
exchange parameter J is approximately $160\:\rm K$, and that the next
nearest neighbor exchange parameter is approximately $0.36\:\rm J$.
\end{abstract}


\newpage

The purpose of the present study is to describe
the spin-Peierls transition in the linear Cu$^{2+}$ spin-1/2 chains
in the inorganic compound CuGeO$_3$ which has been recently
observed\cite{hase,lorenzo,hirota,harris}. The transition temperature
$\rm T_c \approx 14\:K$ has been inferred from the rapid drop of
the magnetic susceptibility towards zero, indicating the opening of
an energy gap for singlet-triplet spin excitations.\cite{nishi,brill}
The existence of this transition was also confirmed by
measurements of the heat capacity\cite{kuroe} which exhibits a
sharp anomaly at $\rm T_c$ corresponding to a second order phase
transition.
We adopt a simple model Hamiltonian consistent of antiferromagnetic
Heisenberg interactions along a chain. Since the one-dimensional (1D)
spin system has no phase transition at finite temperatures, it is
necessary to take into account the spin-lattice coupling in order to
describe the spin-Peierls transition. We consider the spin-lattice
coupling in the adiabatic approximation. The study is performed by
exact diagonalization on finite chains. We suggest that in order
to obtain a reasonable fit of the magnetic susceptibility a next
nearest neighbor antiferromagnetic interaction along a chain should
be included in the model. If the predictions resulting from our model
are confirmed by additional experimental work, the CuO$_2$ chains in
this compound would be one of the few experimental realizations
of a 1D spin-1/2 Heisenberg antiferromagnet
with competing interactions.\cite{bonner85}
The present study should also be considered as part of a current
theoretical effort to understand  magnetic properties, in particular
the singlet-triplet spin gap, in low-dimensional spin systems such
as ${\rm Sr_2Cu_4O_6}$ and ${\rm (VO)_2P_2O_7}$.\cite{barrie}

The thermal properties of spin-Peierls transitions in spin-1/2
antiferromagnetic chains have not been studied, to our knowledge,
from a microscopic and numerical point of view. An analytical study
based on the transformation of Pauli spin operators to spinless
fermion operators has shown that the homogeneous magnetic chain is
unstable with respect to dimerization as the temperature decreases
and a spin gap appears as a result of such
dimerization.\cite{pytte,bulaevsk} The description of experimental
data was usually performed using the mean-field approximation on the
spinless fermion Hamiltonian.\cite{bray83} The effect of a magnetic
field on the spin-Peierls transition temperature has also been
analyzed by Cross and Fisher\cite{crossfisher} using a
Luther-Peschel-type treatment of the spin correlation functions.
The predictions of this theory were compared with experimental
results for CuGeO$_3$,\cite{harris} confirming the
spin-Peierls nature of the observed transition.\cite{bonner87}

The 1D microscopic Hamiltonian for the spin degrees of freedom that
we consider is:
\begin{eqnarray}
\rm H_s = \rm J_1 \sum_{i} {\bf S}_{2i-1} \cdot {\bf S}_{2i}
+ \rm J_2 \sum_{i} {\bf S}_{2i} \cdot {\bf S}_{2i+1}
\label{hamj1j2}
\end{eqnarray}
\noindent
where the index i runs over the lattice cells (i=1,...,N/2, N: number
of sites) with periodic boundary conditions. We assume linear
dependence
of the exchange integrals on the atomic displacements u, so that:
\begin{eqnarray}
\rm J_1 = \rm J\, (1 + \gamma u)     \\
\nonumber
\rm J_2 = \rm J\, (1 - \gamma u)
\end{eqnarray}
\label{j1j2vsu}
\noindent
where $\gamma$ is a constant. It is convenient to introduce the
dimensionless quantity $\delta = \gamma \rm u$.

The spin-lattice interaction enters in this model only through
J$_1$ and J$_2$. The underlying physical picture is the
following.
The spin chains in CuGeO$_3$ are oriented along the c direction
(see for example Fig. 1 in Ref. \cite{hirota}). According to
experimental results\cite{lorenzo} the relevant lattice distortions
are observed along the b axis, perpendicular to the chain direction.
In this situation, the variables u in Eq. (\ref{j1j2vsu})
correspond to oxygen displacements below the spin-Peierls transition
temperature.\cite{harris} It has been recently reported\cite{hirota}
that there is a comparable shift of the Cu ions along the c
direction. The exchange constants J$_1$ and J$_2$, calculated
following the path Cu-O-Cu,
have in principle a complicated dependence on the displacement u.
Then, the expressions for J$_1$ and J$_2$ given above should  be
considered as a first order approximation in u (or $\delta$) of this
complicated function.
Besides, the application of Hamiltonian (\ref{hamj1j2}) to this
material implies that we are neglecting the two-dimensional (2D)
interchain exchange interactions which have a magnitude of
approximately 10 \% of the intrachain coupling.\cite{nishi}

The main difference between the Hamiltonian given by
Eqs. (\ref{hamj1j2}) and (\ref{j1j2vsu}), and the one corresponding
to the dimerized or alternating bond chain,\cite{alterbib,bonner83}
is that in the former
both J$_1$ and J$_2$ are {\em temperature dependent} because
$\delta = \delta(\rm T)$. We calculate the temperature
dependence of these couplings by minimizing
at each temperature the free energy $\cal F$ of the total
Hamiltonian ${\rm H = H_s + H_{ph}}$
with respect to $\delta$ (adiabatic approximation). Then
${\cal F}_{min}(\rm T) = {\cal F}(\delta_{eq}(\rm T))$.
The elastic term of the Hamiltonian is:
\begin{eqnarray}
\rm H_{ph} = \frac{1}{2} {\rm N K u^2}
= \frac{1}{2} \frac{\rm N K}{\gamma^2} \delta^2
\label{elastic}
\end{eqnarray}
where K is the elastic constant. It is also customary to introduce
the dimensionless spin-lattice coupling constant
$\lambda = \rm J \gamma^2 \rm K^{-1}$.
This method of calculation was devised by Beni and Pincus\cite{beni}
for their study of the spin-Peierls transition in a spin chain
with XY interactions. For a finite chain, and for a given set of
parameters {J and $\lambda$}, once we have determined the equilibrium
displacement $\delta_{eq}$ at each temperature, any thermodynamical
quantity can be computed in the dimerized region.

The first stage of our study consisted in estimating the parameters
J and g by fitting the experimental data for the
susceptibility\cite{hase} with the theoretical curve in the uniform
or non dimerized (${\rm J_1 = J_2 = J}$), region. We computed the
susceptibility by generating all energy levels E$_i$ and
their multiplicities $\rm d_i$ in each sector of fixed total S$_z$,
using a Householder algorithm. The susceptibility was then obtained
through its relation to the expected squared magnetization,
summed over energy levels and total $\rm S_z$ sectors;
\begin{eqnarray}
\chi(\rm T) = g^2\mu_B^2\beta\;
{
\sum_{S_z} S_z^2 \; \sum_i \; d_i\; e^{-\beta E_i}
\over
\sum_{S_z} \sum_i \;  d_i\; e^{-\beta E_i} } \ .
\label{suscep}
\end{eqnarray}
\noindent
This approach has the advantage that explicit eigenvectors are not
required.

The observed average g-factor is
approximately 2.14 with a slight anisotropy along the a, b and
c-axis. The experimental data has a broad maximum near 56 K (see
Fig. \ref{fig1}). A fitting of this
data using the uniform Heisenberg model reproduces the position of
this maximum of the susceptibility if the exchange constant J is
chosen to be 88 K.\cite{hase} However, as it can be seen in this
reference the overall fitting is quite poor. A somewhat better
fitting
can be achieved by choosing $\rm {J = 170\:K}$.\cite{hori} However,
for $\rm {T \leq 150\,K}$ the fitting is still quite poor. We have
seen numerically that there are no satisfactory fitting
of the experimental data in the region $\rm T > \rm T_{max} = 56\,K$
using a nearest neighbor Heisenberg model. One of the simplest
extensions of this model is to include in the spin part of the
Hamiltonian a next nearest neighbor interactions term:
\begin{eqnarray}
\rm H_{2n} = \rm J^{\prime} \sum_{j} {\bf S}_j \cdot {\bf S}_{j+2}
\label{2nterm}
\end{eqnarray}
\noindent
where the index j runs over the lattice sites (j=1,...,N). The
possibility of an antiferromagnetic ($\rm J^{\prime} > 0$)
second-neighbor coupling through the Cu-O-O-Cu exchange path was
suggested in Ref. \cite{lorenzo}. Another possibility is to consider
the interchain coupling which would lead to 2D model. Taking into
account the underlying physical picture discussed above, it is
reasonable to assume that $\rm J^{\prime}$ is independent of u at
least in first order approximation. This ${\rm J_1 - J_2 - J^\prime}$
model was studied at zero temperature by Shastry and
Sutherland\cite{suther} for a particular relation of the parameters
where the ground state can be exactly calculated.

We determined J by imposing that the maximum in the susceptibility
is at 56 K as indicated in Ref. \cite{hase}. Then, we determined the
ratio $\alpha_2 = \rm J^{\prime} / \rm J$ in order to fit the
maximum of the susceptibility $\chi_{max} = \chi(\rm T_{max})$.
For this fitting we chose the susceptibility measured in a
polycrystal sample\cite{hori}
shown in Fig. \ref{fig1}. In this figure we also reproduce the
magnetic susceptibility measured on a single crystal\cite{hase}
along the a, b and c axis. Taking into account this dispersion of
the data, we conclude that a reasonable fit is obtained with the
following set of parameters:
\begin{eqnarray}
\nonumber
\rm J = 160\,{\rm K}          \\
\alpha_2 = 0.36
\nonumber
\label{params}
\end{eqnarray}
To simplify the calculations, we have taken g equal to 2.00.
This value of g, which is somewhat smaller than the experimental
value, should be considered as an effective value since we are
neglecting the interchain couplings. Notice that the temperature
region where we are fitting the available experimental data is still
far from the asymptotic regime described by the Curie law. The
effect of the neglected interchain coupling on the fitting parameters
has been discussed in the literature (See e.g. Ref.\cite{bonner83}).
The results for the calculated susceptibility and the experimental
data are shown in Fig. \ref{fig1}. In both theoretical and
experimental data a contribution from the orbital part of the
susceptibility, $\chi^{orb} = 1.5\,10^{-4}\,\,{\rm emu/mole}$ has
been added. In this temperature region, ${\rm T > T_{max}}$,
the finite size effects are negligible and
already for $\rm N = 12$ the results do not
vary by taking larger clusters. For temperatures smaller than
T$_{max}$ but in the uniform region, i.e. above the spin-Peierls
transition temperature, there are strong finite size effects. The
magnitude of these finite size effects can be seen in the inset of
Fig. \ref{fig1} where we show the susceptibility below 30 K for
$\rm N = 8, \,10,\, 12,\, 14$ and 16,
and $\rm N = 9,\, 11,\, 13$ and 15.

A possible unwanted feature of the Heisenberg model with nearest and
next nearest neighbor interactions is the presence of a temperature
independent spin gap, i.e., in the absence of dimerization. This
spin gap can be inferred from an exact solution at zero temperature
found by Majumdar and Ghosh\cite{majumdar} and was confirmed by
subsequent numerical work. Recent studies on this problem indicate
that for $\alpha_2 \geq 0.25$ there is a finite
singlet-triplet gap.\cite{okamoto}
For $\alpha_2 = 0.36$, we have calculated the spin
gap at zero temperature on finite lattices with $\rm N \leq 24$
spins. To extrapolate to the bulk limit we
adopted the form predicted by spin-wave theory,\cite{sano} or
alternatively, the essentially equivalent law,
\begin{eqnarray}
\Delta(\rm N) = \Delta_{\infty} + \frac{\rm c}{\rm N^2}
\label{extralaw2}
\end{eqnarray}
\noindent
The resulting extrapolated value of the spin gap is approximately
equal to $0.015 \pm 0.005$ in units of J, or
$2.4 \pm 0.8 \, \rm K$, much smaller than the smallest measured
value for the CuGeO$_3$.\cite{harris}

The second stage in the calculation was the estimation of the
coupling constant $\lambda$. In order to do this estimation, we chose
another piece of experimental data, the singlet-triplet spin gap at
zero temperature. The fitting of this data is very convenient from
the numerical point of view since at $\rm T = 0$ we can diagonalize
larger lattices using the Lanczos algorithm. Using the spin part of
the Hamiltonian given by Eqs. (\ref{hamj1j2}) and (\ref{2nterm}), we
first
determined the value of $\delta$ that reproduces the experimental
singlet-triplet spin gap which is $\approx 2.15 \,{\rm meV}$
(from Ref. \cite{harris}) or 0.153 in units of ${\rm J = 160\, K}$.
Results for the spin gap for several values of $\delta$ and for
several sizes are shown in Fig. \ref{fig2}a.
The extrapolation of the spin gap to the bulk limit for each value
of $\delta$ was also done using Eq. (\ref{extralaw2}).
The extrapolated spin gaps as a function of $\delta$ are shown in
Fig. \ref{fig2}b. A quadratic interpolation gives the final result:
$\delta_{eq}(\rm T=0) = 0.014 \pm 0.001$.

Then, for each lattice size ${\rm N = 8,10,...,22}$ and for
$\delta = 0.014$, we computed the inverse of the coupling constant
$\lambda(\rm N)^{-1}$ that minimizes the ground state energy of
the total Hamiltonian, ${\rm H_s+H_{2n}+H_{ph}}$. The results are
shown in Fig. \ref{fig2}c.
Finally, by extrapolating $\lambda(\rm N)^{-1}$ to
the bulk limit we obtain $\lambda^{-1} = 21.3 \pm 0.5$ or
$\lambda \approx 0.05$. This value of the spin-lattice coupling is
reasonably small so as to lead to a spin-Peierls transition and not
to a structural one.\cite{bulaevsk}

Now, all the parameters of the Hamiltonian have been determined and
we like to check the validity of this model by reproducing other
experimental results or predicting the value of properties still not
experimentally measured. Alternatively, the values of
$\rm J = 160\, K$ and $\lambda \approx 0.05$ could be determined
independently by other experiments.

The first and most obvious check of the consistency of this model
is the calculation of the spin-Peierls transition temperature
T$_c$. Experimentally, $\rm T_c \approx 14 \,\rm K$.
To estimate T$_c$ we computed the free energies for each lattice
size N, and then, we minimized the free energy to determine
$\delta_{eq}(\rm T,N)$ as explained above. T$_c(\rm N)$ is the value
of the temperature at which the dimerization
begins. For each lattice size,
we used $\lambda(\rm N)^{-1}$ calculated previously. Results for
the calculation of $\delta_{eq}(\rm T,N)$ for $\rm N = 12,\, 14$ and
16 are shown in Fig. \ref{fig3}a. These curves have some resemblance
with the experimental data for the lattice contraction
$\Delta_b$ as shown in Fig. 4 of Ref. \cite{harris}. In fact,
$\Delta_b$ is related to the atomic displacements u of Eq. (2)
and hence to $\delta_{eq}$. The spin-Peierls transition temperature,
calculated for $\rm N = 16$, is approximately $10.5\, \rm K$, which
is reasonably close to the experimental value of $14 \, \rm K$.
We have not yet attempted an extrapolation of $\delta_{eq}(\rm T,N)$
and T$_c(\rm N)$ to the bulk limit.\cite{progress}

As stated above, once $\delta_{eq}(\rm T,N)$ has been computed, any
thermodynamical quantity can be calculated in the dimerized region.
In the first place, we estimated the spin gap as a function of T
for $\rm N = 12,$ 16 and 20. To simplify this calculation we
adopted for the three lattices $\delta_{eq}(\rm T,N)$
corresponding to $\rm N = 16$. At each temperature, we
computed the spin gap as the zero temperature singlet-triplet gap
of the model with $\delta = \delta_{eq}(\rm T,N)$. The results are
shown in Fig. \ref{fig3}b together with the experimental data from
Ref. \cite{harris}. Although there are strong finite size effects
there is a reasonable tendency of the theoretical data towards the
experimental ones.

Finally, we have computed the magnetic susceptibility near the
spin-Peierls transition, for N = 12, 14 and 16. The results are shown
in Fig. \ref{fig4}a,b. In Fig. \ref{fig4}a, it can be seen that the
susceptibility for each lattice size decays as T goes to zero more
rapidly for the dimerized Heisenberg model than for the uniform one,
which is consistent with a larger spin gap in the dimerized case.
Notice that for a {\em finite} chain there is always a finite gap
even in the absence of dimerization and for $\rm J^{\prime} = 0$.
As the lattice size is increased, the spin gap in the dimerized
model remains finite while the spin gap of the uniform model
drops to a small value (due to the presence of $\rm J^{\prime}$).
This behavior explains the fact that the difference between the
dimerized and the uniform curves is larger as the lattice size
is increased, as it can be seen in this figure.

In
Fig. \ref{fig4}b, we compare the theoretical susceptibility obtained
for the 16 site chain with experimental data from Ref. \cite{hase}
obtained with a small magnetic field parallel to the c axis.
The spin-Peierls transition temperature calculated previously, is
shown with an arrow. The agreement between theoretical and
experimental results is quite good taking into account the
approximations involved and the strong finite size effects expected
in this low temperature region.

In summary, we described the magnetic susceptibility, the
temperature dependence of the spin gap and the spin-Peierls
transition
temperature of the CuGeO$_3$ using a one-dimensional
antiferromagnetic Heisenberg model with nearest and next nearest
interactions. We obtained a quite satisfactory overall agreement with
experimental results with only three free parameters. This agreement
gives in turn support to the interpretation of the observed
features as a spin-Peierls transition.
The nearest
neighbor exchange is approximately equal to $160 \, \rm K$,
and the ratio of next nearest to nearest neighbor exchange constants
is approximately 0.36. This value of the parameter $\alpha_2$ would
imply a spin gap of the order of $2.4 \, \rm K$, even in
the absence of dimerization. If this spin gap is not detected
experimentally, it is quite apparent that one should necessarily
adopt a two-dimensional model.
In fact, some small discrepancies between our model
and experiment should be attributable to weak interchain coupling
as well to a slight spin anisotropy. However, we don't think that
it is
relevant to include these effects before confirmation of the main
consequences derived from the present model.
Details of the calculations and a systematic study of finite size
effects are discussed in an enlarged version
of this report.\cite{progress}

We acknowledge many useful discussions with A. Greco and S. Koval,
especially during the early stage of this study, and with R. Calvo.

\newpage

\figure{ The magnetic susceptibility of CuGeO$_3$. Experimental
curves labelled a, b and c, obtained on a single crystal, are from
Ref. \cite{hase}. Experimental curve with dashed line corresponds
to measurements on a polycrystal.\cite{hori} The solid curve
is a theoretical one corresponding to the Heisenberg model
with nearest and next nearest neighbor interactions, with
$\alpha_2 = 0.36$, obtained numerically on a chain with 16 sites.
In the inset, the theoretical susceptibility obtained for lattices
with even and odd number of spins for ${\rm T < 30\,K}$ are shown.
\label{fig1}}

\figure{a) Spin gap (in units of J), at $\rm T = 0$, for the
dimerized Heisenberg model with nearest and next nearest neighbor
interactions as a function of the inverse of the lattice size and
for several values of $\delta$.
b) Spin gap in units of J, at $\rm T = 0$, as a function of $\delta$
in the bulk limit. The diamond indicates the point corresponding to
the experimental value of the spin gap.
c) Inverse of the adimensional spin-lattice coupling constant
$\lambda$ as a function of the inverse of the lattice size.
The dashed line corresponds to a quadratic extrapolation which
leads to $\lambda^{-1} \approx 21.3$ in the bulk limit.
\label{fig2}}

\figure{a) $\delta_{eq}$ (see text) as a function of the temperature
for N = 12, 14 and 16.
b) Spin gap as a function of the temperature for N = 12, 16 and 20.
Experimental data from Ref. \cite{harris} are indicated with solid
circles.
\label{fig3}}

\figure{a) Magnetic susceptibility, in arbitrary units, near the
spin-Peierls transition for N= 12, 14, and 16 as a function of the
temperature. The solid (dashed) curves correspond to the
non-dimerized (dimerized) Heisenberg chains.
b) Comparison of the theoretical susceptibility obtained for the 16
site chain with experimental data from Ref. \cite{hase} obtained with
a magnetic field parallel to the c axis.
\label{fig4}
}

\end{document}